\documentclass[fleqn,usenatbib]{mnras}

\usepackage{newtxtext,newtxmath}
\usepackage{xcolor}
\usepackage[T1]{fontenc}

\DeclareRobustCommand{\VAN}[3]{#2}
\let\VANthebibliography\thebibliography
\def\thebibliography{\DeclareRobustCommand{\VAN}[3]{##3}\VANthebibliography}

\usepackage{graphicx,xcolor,enumitem,academicons,xspace,color,longtable}	% Including figure files 
\usepackage{amsmath,shadow,bezier,curves,rotating}	% Extra maths symbols
\usepackage{cleveref}
\usepackage[flushleft]{threeparttable}

\newcommand{\Msun}{M_{\odot}}
\newcommand{\hinv}{h^{-1}}

\def\r200m{r_{\rm 200m}}
\def\Gpc{{\rm Gpc}}

\def\Msun{{\rm M}_\odot}
\def\kpc{{\rm kpc}}
\def\kms{{\rm km\,s}^{-1}}

\newcommand{\urluchuu}{\url{http://www.skiesanduniverses.org/}}
\newcommand{\SU}{{Skies \& Universes}}
\newcommand{\UM}{\textsc{UniverseMachine} }

\title[Uchuu-UM Galaxy catalogues]
{The Uchuu-UniverseMachine dataset: Galaxies in and around Clusters}

\author[Aung et al.]{Han Aung$^{1,2}$\thanks{han.aung@mail.huji.ac.il},
Daisuke Nagai$^{1}$, 
Anatoly Klypin$^{3}$,
Peter Behroozi$^{4}$,
Mohamed H. Abdullah$^{5,6}$,
\newauthor
Tomoaki Ishiyama$^{5}$,
Francisco Prada$^{7}$, 
Enrique P\'erez$^{7}$, 
Javier~L\'opez Cacheiro$^{8}$,
Jos\'e Ruedas$^{7}$
\\
$^{1}$Department of Physics, Yale University, New Haven, CT 06520, USA\\
$^{2}$Centre for Astrophysics and Planetary Science, Racah Institute of Physics, The Hebrew University, Jerusalem 91904, Israel\\
$^{3}$Department of Astronomy, New Mexico State University, Las Cruces, 88003 NM, USA\\
$^{4}$Steward Observatory, University of Arizona, Tucson, 85719 AZ, USA\\
$^{5}$Institute of Management and Information Technologies, Chiba University, 1-33, Yayoi-cho, Inage-ku, Chiba, 263-8522, Japan\\
$^{6}$Department of Astronomy, National Research Institute of Astronomy and Geophysics, Cairo, 11421, Egypt\\
$^{7}$Instituto de Astrof\'isica de Andaluc\'ia (CSIC), Glorieta de la Astronom\'ia, E-18080 Granada, Spain\\
$^{8}$Centro de Supercomputaci\'on de Galicia (CESGA), Avenida de Vigo, s/n Campus Sur, E-15705 Santiago de Compostela, Spain
}

\date{Accepted XXX. Received YYY; in original form ZZZ}

\pubyear{2022}

\hypersetup{draft}

\begin{document}
\label{firstpage}
\pagerange{\pageref{firstpage}--\pageref{lastpage}}
\maketitle

\begin{abstract}
We present the public data release of the Uchuu-UM galaxy catalogues by applying the \UM algorithm to assign galaxies to the dark matter halos in the Uchuu $N$-body cosmological simulation. It includes a variety of baryonic properties for all galaxies down to $\sim 5\times10^8 M_{\odot}$ with halos in a mass range of $10^{10}<M_{\rm halo}/M_{\odot}<5\times10^{15}$ up to redshift $z=10$. Uchuu-UM includes more than $10^{4}$ cluster-size halos in a volume of $ 8(\hinv \Gpc)^3$, reproducing observed stellar mass functions across the redshift range of $z=0-7$, galaxy quenched fractions, and clustering statistics at low redshifts. 
Compared to the previous largest UM catalogue, the Uchuu-UM catalogue includes significantly more massive galaxies hosted by large-mass dark matter halos. Overall, the number density profile of galaxies in dark matter halos follows the dark matter profile, with the profile becoming steeper around the splashback radius and flattening at larger radii. The number density profile of galaxies tends to be steeper for larger stellar masses and depends on the color of galaxies, with red galaxies having steeper slopes at all radii than blue galaxies. The quenched fraction exhibits a strong dependence on the stellar mass and increases toward the inner regions of clusters.
The publicly available Uchuu-UM galaxy catalogue presented here can serve to model ongoing and upcoming large galaxy surveys.
\end{abstract}


%%%%%%%%%%%%%%%%%%%%%%%%%%%%%%%%%%%%%%%%%%%%%%%%%%

%%%%%%%%%%%%%%%%% BODY OF PAPER %%%%%%%%%%%%%%%%%%

\section{Introduction}

In the era of large galaxy surveys, one of the key challenges lies in creating an observable universe or mock galaxy catalogue with a very large number of galaxies to constrain statistical uncertainties, with correct properties of galaxies from sophisticated non-linear astrophysics. Empirical models that employ a galaxy-halo connection to add galaxies into halos and subhalos of N-body simulations provide a unique opportunity for generating a mock catalogue large enough to produce meaningful statistics, while reproducing observed galaxy properties. 

The combination of extensive $N$-body simulations such as MultiDark Planck \citep[MDPL2,][]{multidark} and the \UM galaxy model \citep{Behroozi19} has emerged as a powerful theoretical tool for modeling large galaxy surveys. \UM provides a computationally efficient approach to generate significant statistics of galaxies by pasting galaxies onto the halos and subhalos in the dark matter only $N$-body simulations. \UM models galaxy formation and evolution by relating the dark matter halo's maximum circular velocity and accretion history to the star formation rate of the galaxy. It can model the halo-galaxy connection over a broad range of cosmic time and environments and reproduces a plethora of observational datasets on galaxy formation and evolution \citep{Behroozi19,Behroozi20}. For example, to date, the MDPL2-UM galaxy catalogue has been used to investigate stellar mass-halo mass relations \citep{Bradshaw2020}, galaxy formation physics \citep{Montero2021}, the phase space structure of dark matter halos \citep{Aung2022}, and dynamical mass measurements of galaxy clusters \citep{Ho2019} among others. 

The  Uchuu $N$-body simulation \citep{Uchuu} provides a substantial statistical sample of dark matter halos with 8 times larger volume and 5 times better mass resolution than the previous largest simulation combined with \UM{} (MDPL2). The Uchuu simulation allows us to capture and resolve the structure and concentration of dark matter halos across cosmic time ($0\leq z \leq 14$) \citep{Uchuu}, as well as the subhalo properties to as close as $\sim$0.1 percent of the host virial radius \citep{Moline2021}. Thus, the Uchuu simulation, combined with the \UM algorithm, provides the required statistics and resolution for the next-generation galaxy surveys.

In this work, we present and analyze an Uchuu-UM galaxy catalogue in $8(\hinv {\rm Gpc})^3$, encompassing over $10^4$ cluster-size halos of mass $>10^{14}\Msun$ while resolving galaxies down to $M_*\sim 5\times 10^8\Msun$. We show that the catalogue reproduces the observed galaxy clustering and quenched fractions. We show how the number density profiles and quenched fractions of galaxies in the cluster and group-size dark matter halos depend on the galaxy properties, such as stellar mass and colors.

The paper is organized as follows. We describe the Uchuu simulation and UM galaxy model and present the Uchuu-UM galaxy catalogue and data format in Section~\ref{sec:method}. We present the properties of galaxies in the Uchuu-UM catalogue in Section~\ref{sec:properties}. We summarize our main conclusion in Section~\ref{sec:conc}.

\section{Uchuu-UM Galaxy catalogue Data} 
\label{sec:method}

\subsection{Simulation}
Uchuu simulation has a comoving box-size of $2\hinv\Gpc$ with a total of $12800^3\sim $ 2.1 trillion particles. The simulation was run using the massively parallel TreePM code
\textsc{GreeM} \citep{Ishiyama09,Ishiyama12} for the Planck (2018) cosmology: $h  = 0.6774$, $\Omega_{\rm m} = 0.3089$, $\Omega_{\rm b} = 0.0486$, $n_{\rm s} = 0.9667$, $\lambda_0=0.6911$, and $\sigma_8 = 0.8159$ \citep{Planck18_cosmology}, resulting in a dark matter particle mass resolution of $3.27\times 10^8 \hinv\Msun$ and a gravitational softening length of $4.27 \hinv\kpc$. 
The halo catalogue is generated using the \textsc{Rockstar} code \citep{Behroozi13Roc}, and the merger tree is generated by the \textsc{ConsistentTrees} algorithm \citep{Behroozi13Con}. Full simulation details can be found in \citet{Uchuu}.

\subsection{UniverseMachine}
The mock galaxy catalogue is constructed using the \UM  \citep{Behroozi19},\, in which the galaxy--halo relationship is forward modelled to match observational data across cosmic times.
The in-situ star formation rate is parameterized as a function of halo mass, halo assembly history, and redshift. We use this relation to assign the star formation rate for each halo and subhalo in the Uchuu simulation. 
The stellar mass of the galaxy hosted by the halo is then computed by integrating the star formation rates over the merger history of the halo, accounting for mass lost during stellar evolution. The parameters of the algorithm were explored by a Markov Chain Monte Carlo (MCMC) algorithm, and a fitting algorithm optimized the best-fitting parameters so that the resulting mock catalogue from the \UM Data Release 1 reproduces the following observational data across a broad range of redshifts ($0<z<10$): (i) stellar mass functions; (ii) cosmic star-formation rates and specific star-formation rates; (iii) UV luminosity functions; (iv) quenched fractions; (v) correlation functions for all, quenched, and star-forming galaxies; (v) measurements of the environmental dependence of central galaxy quenching; and (vi) UV-stellar mass relations. These comparisons are based on the data obtained from the Sloan Digital Sky Survey, the PRIsm MUlti-object Survey, and the CANDELS survey, among others \citep[for more details, see the Appendix in ][]{Behroozi19}. 

To compensate for artificially disrupted subhalos in the simulations, \UM adds ``orphan'' galaxies, which do not correspond to any subhalos or halos in the halo catalog of the corresponding snapshot, but are descendants of the subhalos in previous snapshots. Orphans are added by leapfrog integration of the positions and velocities of disrupted subhalos, with a modified mass and $v_{\rm max}$ loss prescription based on \citet{jiang_vdB14}. We retain orphans until $v_{\rm max}$ of the subhalo drops below a mass-dependent threshold, which is fitted with two free parameters.

\begin{table*}
    \centering
    \begin{tabular}{p{2.5cm}|p{2cm}|p{1.5cm}|p{10cm}}
    \hline
    parameter & units & data type & description\\
    \hline
    \hline
    id   & n/a & int64 & ID of the galaxy. Share the ID with dark matter halo  (appeared in \textsc{Rockstar} catalogs and merger trees) in which galaxy resides, if it is not an orphan.  For orphans, the id is the last ID in the halo catalog + $10^{15} \times$ (number of snapshots as an orphan)\\
    upid  &  n/a & int64 & ID of the largest dark matter halo in which galaxy resides ($-1$ for central halos)\\
    Mvir & $\Msun$ & float32 & virial mass~\citep{Bryan1998} of the dark matter halo \\
    sm & $\Msun$ & float32 & true stellar mass of the galaxy\\
    icl & $\Msun$ & float32 & true intracluster stellar mass inside the dark matter halo\\
    sfr & $\Msun{\rm yr}^{-1}$ & float32 & true star formation rate of the galaxy\\
    obs\_sm & $\Msun$ & float32 & observed stellar mass of the galaxy including random and systematic errors\\
    obs\_sfr & $\Msun{\rm yr}^{-1}$ & float32 & observed star formation rate of the galaxy including random and systematic errors\\
    obs\_um & M1500,AB & float32 & observed UV Magnitude \\
    \hline
    x, y, z & comoving ${\rm Mpc}\,h^{-1}$  & float32 & comoving x,y,z position of the galaxy\\
    \hline
    desc\_id & n/a & int64 & ID of descendant halo (or -1 at $z=0$)\\
    vx, vy, vz & peculiar ${\rm km}\,s^{-1}$ & float32 & peculiar velocity of the galaxy\\
    Mpeak & $\Msun$ & float32 & halo peak historical mass (virial mass)\\
    Vmax\_Mpeak &  ${\rm km}\,s^{-1}$ & float32 & maximum circular velocity when peak mass was achieved\\
    vmax &  ${\rm km}\,s^{-1}$ & float32 & maximum circular velocity of the dark matter halo \\
    A\_UV & mag & float32 & UV attenuation \\
    \hline
    SFH & $\Msun{\rm yr}^{-1}$ & float32$\times N$ & star formation rate history for the present-day stellar population in the galaxy (including all merged progenitors)\\
    ICLH & $\Msun{\rm yr}^{-1}$ & float32$\times N$ & star formation rate history for the present-day population in the intrahalo light\\
    SM\_main\_progenitor & $\Msun$ & float32$\times N$ & the main progenitor galaxy's stellar mass history\\
    ICL\_main\_progenitor & $\Msun$ & float32$\times N$ & the main progenitor's intrahalo light history\\
    M\_main\_progenitor & $\Msun$ & float32$\times N$ & the main progenitor's halo mass history\\
    SFR\_main\_progenitor & $\Msun{\rm yr}^{-1}$ & float32$\times N$ & the main progenitor's star formation rate history (excluding any mergers)\\
    V@Mpeak & ${\rm km}\,s^{-1}$ & float32$\times N$ &  the main progenitor's Vmax\_Mpeak history\\
    A\_first\_infall & n/a & float32 & Scale factor at which the galaxy first passed through a larger halo\\
    A\_last\_infall & n/a & float32 & Scale factor at which the galaxy last passed through a larger halo\\
    \hline
    \end{tabular}
    \caption{The columns in the dataset for the Uchuu-UM catalogues, where each column is stored in HDF5 files, with a given data type. $N$ is the number of snapshots before the current snapshot (e.g., $N=50$ at $z=0)$.}
    \label{tab:data}
\end{table*}

\Cref{tab:data} describes the data presented in the Uchuu-UM catalog, generated from 50 snapshots of the halo catalogs spanning $z=14$ to $z=0$. We report the halo mass and radius of the spherical overdensity  halo mass definition $M_{\rm halo} = M_{\rm 200m}$ provided by the Rockstar algorithm.\footnote{We note that the outer profiles of the halos are self-similar when the radius is normalized using $r_{\rm 200m}$, i.e., the profiles are independent of mass and redshift\citep{diemer_kravtsov2014,more_etal2015}, while the inner profiles are self-similar when normalized using $r_{\rm 200c}$ \citep{Lau2015}.} 
For the subhalo masses, we use $M_{\rm sub}=M_{\rm vir}$ because the \UM algorithm provides only the virial masses of the orphan galaxies. For galaxy stellar masses, we use the true stellar mass provided by \UM in units of $M_{\odot}$.

\subsection{Comparison with observations}

Here, we show how well the Uchuu-UM catalogue reproduces the observed data of galaxy clustering and quenched fractions that are used to constrain the parameters of the \UM algorithm. We define galaxies that have specific star formation rates of ${\rm sSFR}>10^{-11}\Msun {\rm yr}^{-1}$ as galaxies that are star-forming (blue), and those with ${\rm sSFR}<10^{-11}\Msun {\rm yr}^{-1}$ as quenched galaxies (red).

\begin{figure}
    \centering
    \includegraphics[width=0.49\textwidth]{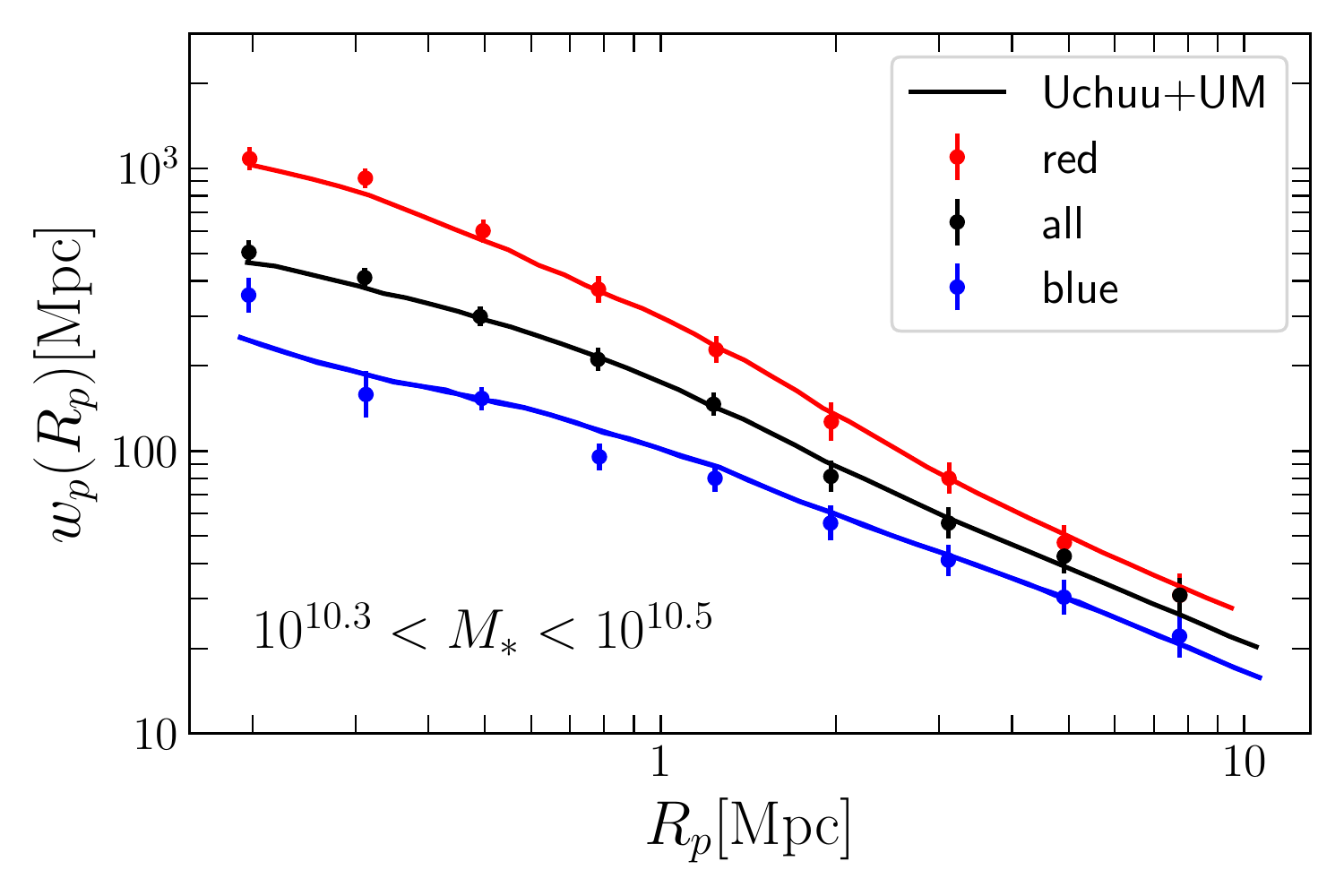}\\
    \includegraphics[width=0.49\textwidth]{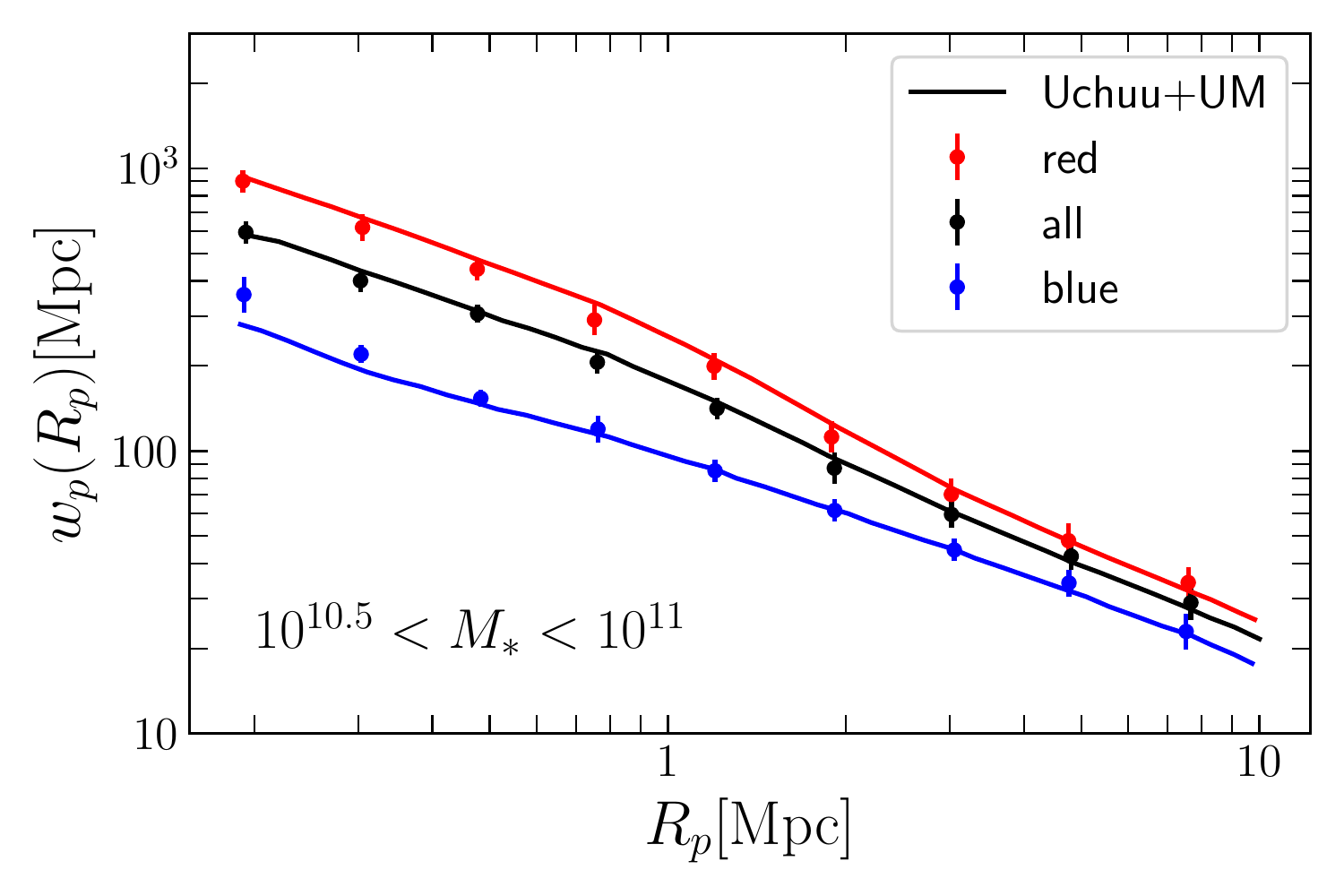}
    \caption{Comparison of projected correlation functions of mock galaxies (full curves) with observations (symbols; \citet{Behroozi19}. The correlation function is more significant for larger stellar mass galaxies or quenched galaxies as they preferentially reside in dense environments. Stellar masses are given in units of $M_{\odot}$.}
    \label{fig:correlation}
\end{figure}

\Cref{fig:correlation} shows the projected correlation function of red, blue, and all galaxies as a function of projected radius for different stellar mass bins. The correlation function measured in the simulation agrees well with the observed data compiled in \citet{Behroozi19}. The correlation function increases for larger stellar mass as the brighter, more massive galaxies reside in larger mass halos which cluster more strongly\citep{Scranton2002,Connolly2002}. Red, quenched galaxies also exhibit a stronger correlation than blue, star-forming galaxies because the quenching of satellite galaxies, which produces red galaxies, happens preferentially in clusters and groups, which are more clustered  \citep[e.g.,][]{Hearin2013,Hearin2014,Watson2015}. 

\begin{figure}
    \centering
    \includegraphics[width=0.49\textwidth]{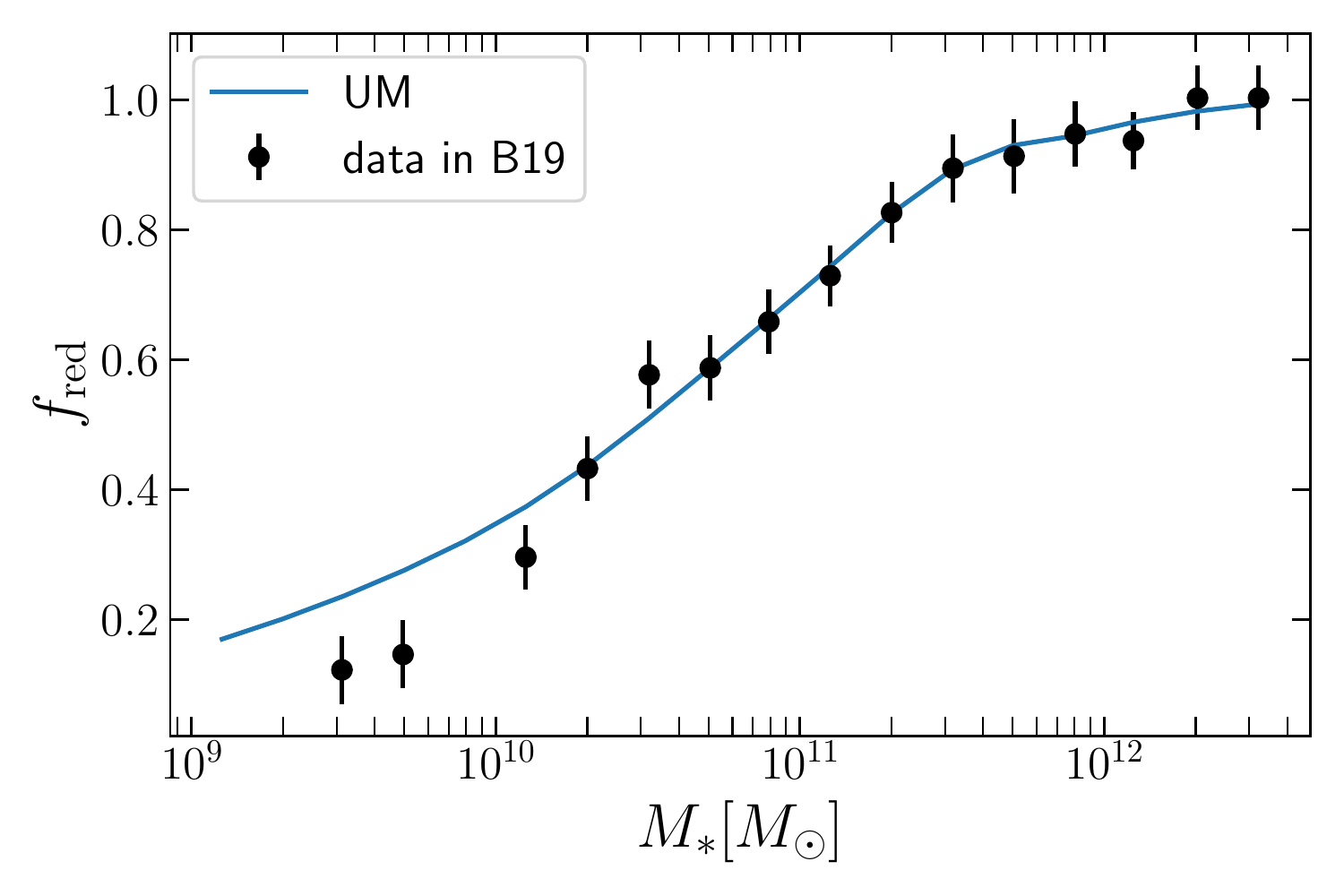}
    \caption{Quenched fraction of galaxies as a function of stellar mass in comparison to the observational data compiled in \citet{Behroozi19} at $z=0.1$. Galaxies at large stellar masses are quenched due to stellar and AGN feedback, which are less effective at small stellar masses. The dataset provides essential data to constrain the quenched fraction for large mass galaxies at $M_*>10^{12}\Msun$.}
    \label{fig:quenched_fraction}
\end{figure}

\Cref{fig:quenched_fraction} shows that the global quenched fraction as a function of the galaxy's stellar mass agrees with the observed quenched fraction compiled in \citet{Behroozi19}. We define the quenched fraction $f_{\rm red}$ as the ratio of red galaxies to the total number of galaxies. This figure demonstrates that galaxies of smaller mass are more star-forming, while those with large mass are more quenched. Even though our simulation does not include baryonic physics, the dependence of the quenched fraction on stellar mass is reproduced well. 

\section{Statistics and properties of galaxies in the Uchuu-UM catalogue}
\label{sec:properties}

In this section, we study the statistics and properties of galaxies in the Uchuu-UM catalogue, focusing on the stellar mass function, radial distribution of galaxies, and quenched fraction. Because of the unprecedented volume and high resolution of the Uchuu simulation, the \UM mock catalogue allows us to probe galaxies with stellar masses as large as  $\sim 10^{12}\Msun$ while including galaxies down to $10^{8}\Msun$. This also allows us to measure the radial distributions of galaxies as a function of halo radius for galaxy clusters and groups for dynamic mass ranges of satellite galaxies, where the addition of orphan galaxies mitigates artificial disruption of subhalos.

\subsection{Stellar Mass Functions}
\label{sec:smf}

\begin{figure}
    \centering
    \includegraphics[width=0.47\textwidth]{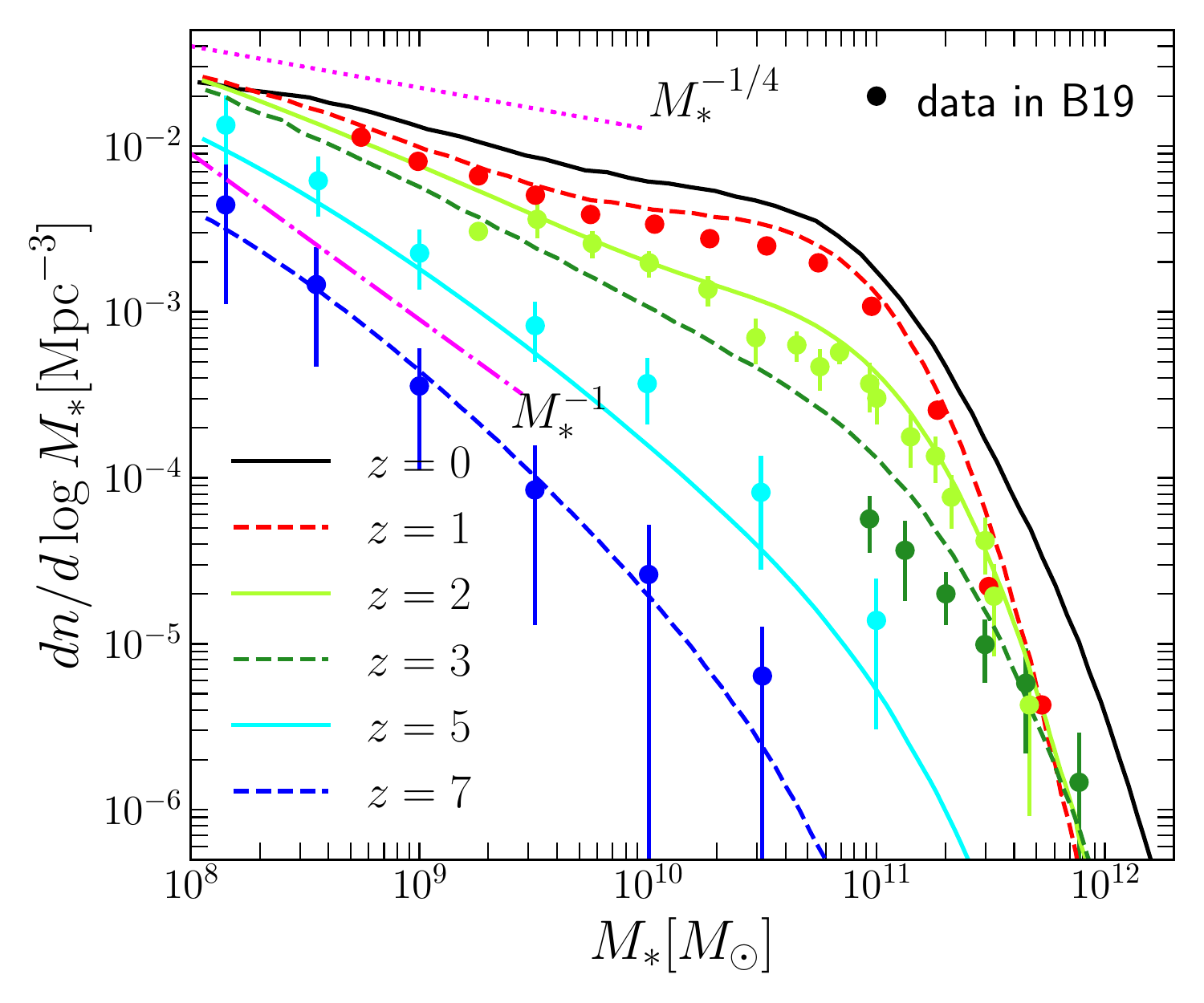}
    \caption{Evolution of the stellar mass function (SMF) between $z=0$ and $z=7$. The observed data compiled in \citet{Behroozi19} is shown with errorbars. At small masses $M_*< 10^{10}\Msun$ SMF is steep at high redshifts and gradually becomes shallower at low $z$. To a large degree, this is driven by the fast growth of stellar mass in Milky-Way-type galaxies with halo masses $M_{\rm halo}\sim 10^{12}\Msun$ and stellar masses $M_*\sim (3\times 10^{10}-10^{11})\Msun$. Because of the high efficiency of star formation in these galaxies, the slope of SMF at larger masses dramatically steepens over time. Evolution from $z=1$ to $z=0$ of the high-mass tail of the SMF $(M_*>10^{11}\Msun)$ is not related to star formation: galaxies with these masses are quenched and do not form new stars. However, they grow substantially by tidally stripping galaxies that move close to them. }
    \label{fig:smf}
\end{figure}

\begin{figure}
    \centering
    \includegraphics[width=0.49\textwidth]{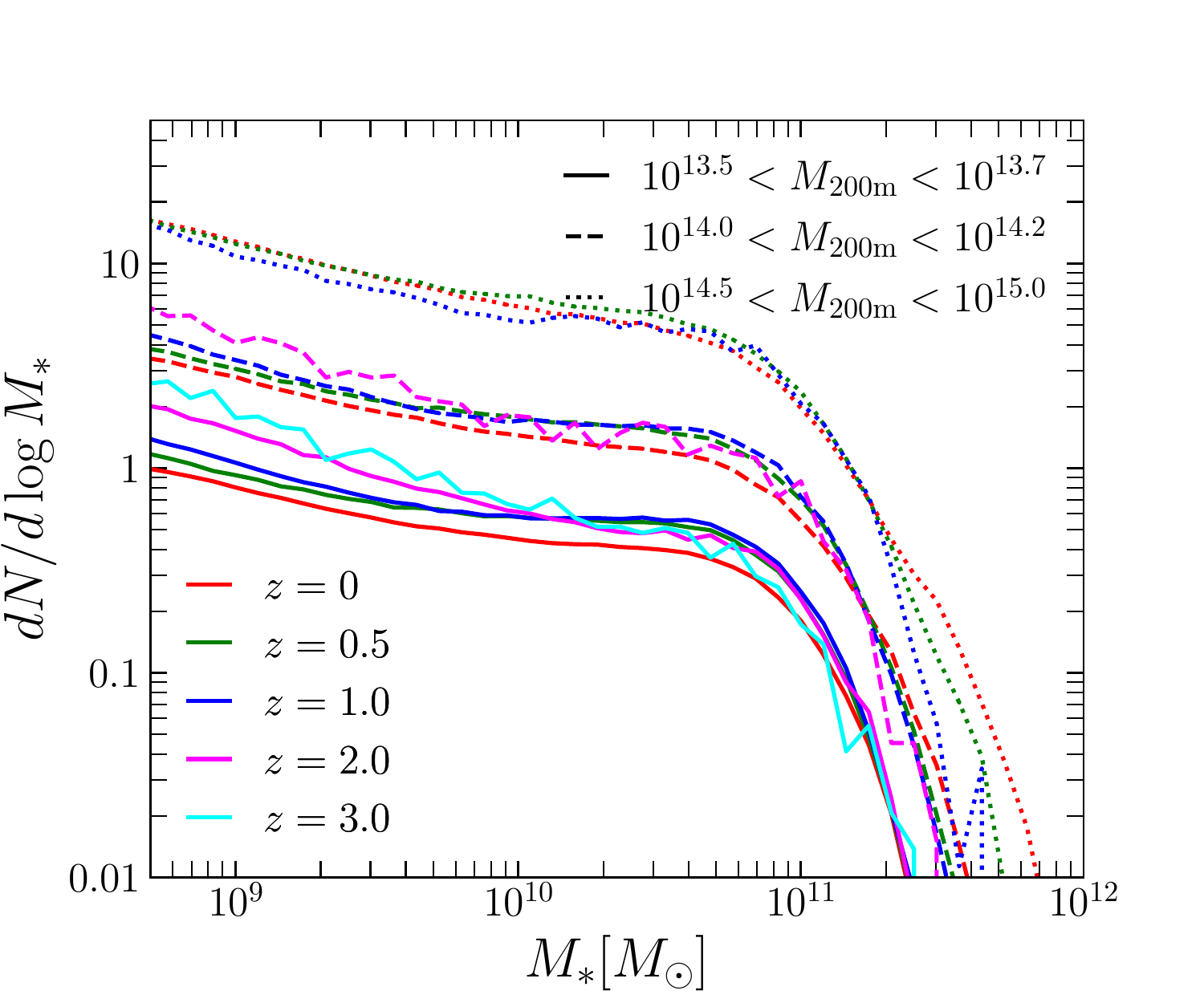}
    \caption{Satellite stellar mass function: the number of satellites per dark matter halo for different dark matter halo masses for the redshift range of $0<z<3$. As expected, the number of satellites increases monotonically with halo mass and decreases with redshift (for the same halo mass).}
    \label{fig:csmf}
\end{figure}

The stellar mass function (SMF) of the Uchuu-UM sample is similar to the SMF based on the Bolshoi-Planck simulation.\footnote{Note that the cosmological parameters of these simulations were not identical to those of Uchuu.} 
The main difference lies in the size of the simulation volume, where the Uchuu simulation has a volume 512 larger than that of the Bolshoi-Planck simulation. The stellar mass function is complete above $\sim 5\times10^8\Msun$ at $z=0$, as the stellar mass function starts to flatten below this mass. This lower bound stellar mass for the completeness limit is smaller at higher redshifts. 

\Cref{fig:smf} shows the evolution of the stellar mass function.
At small masses ($M_*< 10^{10}\Msun$), the SMF becomes steeper at higher redshifts and gradually becomes shallower at low-$z$. To a large degree, this is driven by the fast growth of stellar mass in Milky-Way-type galaxies with halo masses $M_{\rm halo}\sim 10^{12}\Msun$ and stellar masses $M_*\sim (3\times 10^{10}-10^{11})\Msun$.

The SMF's shape and evolution at $z>3$ is qualitatively similar to the evolution of the halo mass function. At small masses, the slope of the stellar mass function $dn/dM$ is steeper with redshift, the same as for the halo mass function, and approaches $-2$ (dot-dashed line in the plot). At larger masses, the SMF declines, but the decline is gradual -- compare $z=5$ and $z=1$ curves -- similar to the halo mass function. The 'knee' -- a point of transition from the small-scale power-law trend to a steeper decline -- increases in mass as the Universe expands.

The evolution of the SMF starts to change at $z<3$.
Because of the high efficiency of star formation in $M_*\sim (3\times 10^{10}-10^{11})\Msun$ galaxies, the slope of SMF at larger masses dramatically steepens over time. 
Evolution from $z=1$ to $z=0$ of the high-mass tail of SMF $(M_*>10^{11}\Msun)$ is not related to star formation: galaxies with these masses are quenched and do not form new stars. However, just as observed galaxies \citep[e.g.,][]{vanDokkum2010,vanderWel2014}, they grow substantially by minor mergers and tidally stripping galaxies that move close to them \citep[e.g.,][]{Naab2009,Damjanov2022}.

\Cref{fig:csmf} shows results on the satellite stellar mass function: the number of satellites in halos with different dark matter masses. As expected, the number of satellites of nearly all masses roughly increases linearly with the halo mass, and the increase in the satellite abundance does not depend on the satellite mass $M_*$. The satellite abundance also increases as the redshift of the halo increases. Dark matter halos of the same mass at earlier redshift have higher rates of satellite accretion and hence, larger numbers of satellites \citep{vdb2005}. However, the trend with redshift disappears for the largest mass cluster bin, where the clusters are actively accreting even at redshift $z=0$. However, note that the lack of redshift evolution in the satellite stellar mass function at the highest mass bin does not imply the lack of evolution for individual halos, as the halos are growing in mass over time.

$N$-body simulations tend to suffer from numerical effects, especially on small scales, due to limited particle count and force resolution. Specifically, the properties of halos and subhalos (including $v_{\rm max}$ with smaller masses) may not be estimated correctly \citep{Mansfield2021}. As our galaxy catalogue places galaxies directly onto subhalos identified from \textsc{Rockstar} in the simulation and orphans extrapolated by our algorithm, the validity of the galaxy properties depends on the $v_{\rm max}$ values in the simulation. The ratio of maximum circular velocity to the circular velocity at $r_{\rm 200m}$ ($v_{\rm max}/v_{\rm 200m}$) in our simulation decreases with increasing mass above a certain halo mass limit $\sim 2\times 10^{11} h^{-1}\Msun$. Below this limit, the ratio increases with increasing mass, indicating a resolution limit \citep[see][]{Mansfield2021}.
Notably, this corresponds to a stellar mass of $\sim5\times 10^{8}\Msun$ in the stellar mass-halo mass relations. Thus, we caution the readers to treat galaxy properties below this mass threshold with care.

\subsection{Number-density profiles}
\label{sec:radprof}

\begin{figure}
    \centering
    \includegraphics[width=0.49\textwidth]{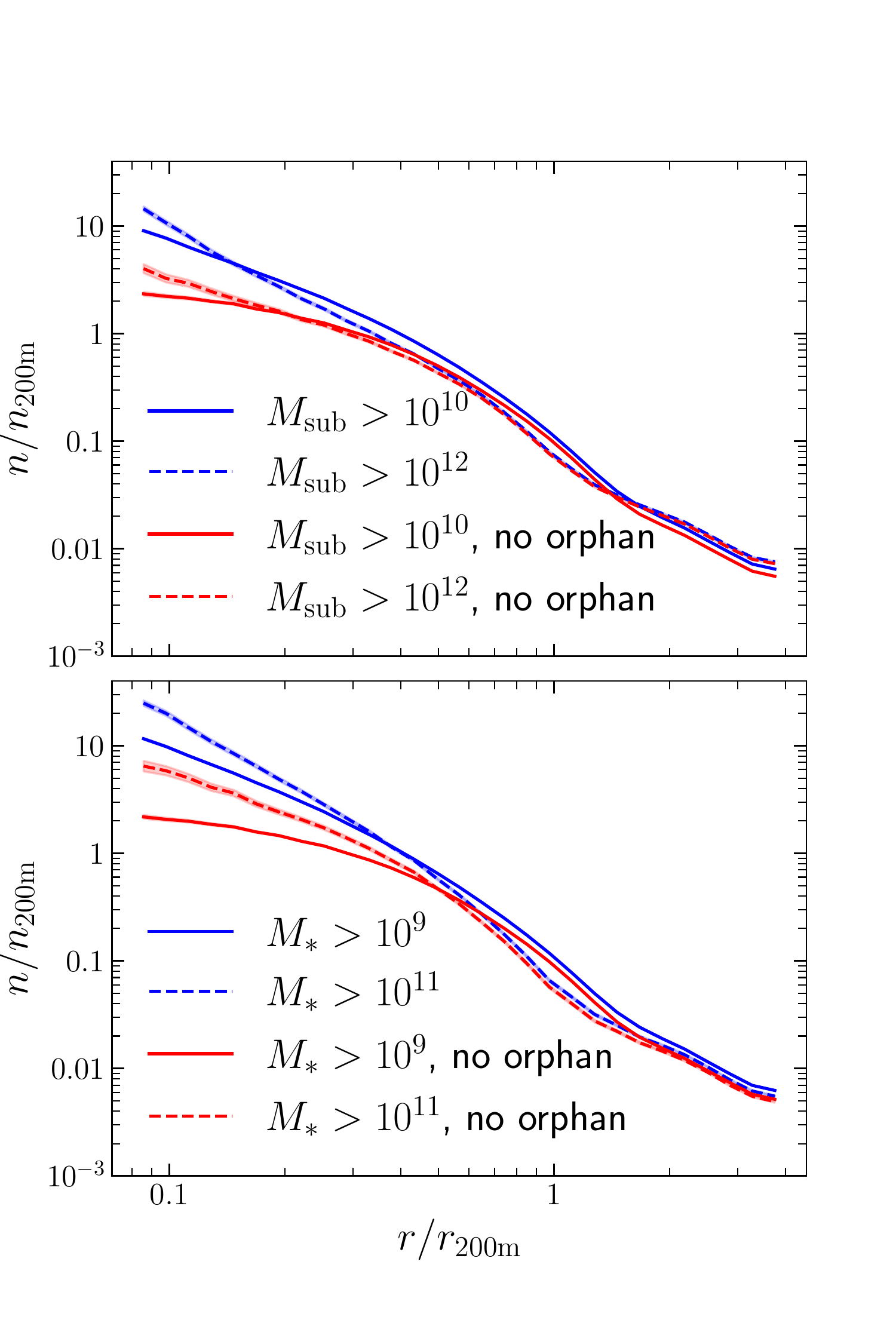}
    \caption{The 3D number-density profile of galaxies as a function of the normalized radius, $r/r_{\rm 200m}$, for halos of $10^{14}h^{-1}M_\odot<M_{\rm 200m}<10^{15}h^{-1}M_\odot$ at $z=0$. The number-density is normalized by the mean number-density of galaxies enclosed within $r_{\rm 200m}$, $n_{\rm 200m} = N_{\rm 200m}/V_{\rm 200m}$, where $V_{\rm 200m}$ is the volume enclosed and $N_{\rm 200m}$ is the total number of galaxies within $r_{\rm 200m}$ with the stellar mass or subhalo mass cut including orphans. {\it Top panel:} the profiles for different subhalo mass cut, $M_{\rm sub}$, in the units of $h^{-1}M_{\odot}$. {\it Bottom panel:} the profiles for different satellite stellar mass cuts, $M_{\ast}$, in the units of $M_{\odot}$. Both panels show the profiles with (solid lines) and without (dashed lines) orphan galaxies. Profiles with different subhalo or stellar mass cuts are shown with different colors. Inclusion of orphan galaxies changes the inner $r< 0.3r_{\rm 200m}$ number-density profile, but not the outer region.}
    \label{fig:number_density}
\end{figure}

\Cref{fig:number_density} shows the  average number-density profile of galaxies as a function of normalized radius, $r/r_{\rm 200m}$ for halos with masses $10^{14}h^{-1}M_\odot<M_{\rm 200m}<10^{15}h^{-1}M_\odot$.
Galaxies were selected either by the dark matter mass (top panel) or stellar mass (bottom panel). The shaded areas in the plot indicate the errors of the mean. Inclusion of orphan galaxies 
changes the inner $r< 0.3\,r_{\rm 200m}$ region of the profile: the slope of the profile increases when orphan galaxies are included. This happens regardless of galaxy selection, either by stellar or subhalo mass. Note also that the profile gets steeper for more massive galaxies.  

The situation is different in the outer regions of the halos. Here the number-density profiles do not depend on how galaxies are selected. 
Suppose we ignore orphan galaxies added by the \UM algorithm. In that case, the profile starts to flatten in the inner region instead of maintaining the cuspy shape that dark matter density profiles exhibit. Orphans are added instead of subhalos which cannot be tracked by the halo finder anymore as they are artificially disrupted. 

However, the decrease in slope/plateauing in the inner regions of the clusters with or without orphan galaxies is much smaller than previously reported by \citet{Nagai05}. The difference likely stems from the halo finder Bound Density Maxima, \textsc{BDM} employed by \citet{Nagai05}. \textsc{BDM} does not give an accurate estimate of subhalo mass and tends to miss subhalos in the inner region of the clusters and groups \citep{Knebe11}. The number-density profiles in the inner part of the clusters without orphan galaxies likely depend on the halo finder employed, as we find that the slope of the number-density profile increases with increasing subhalo or stellar mass.

\begin{figure}
    \centering
    \includegraphics[width=0.49\textwidth]{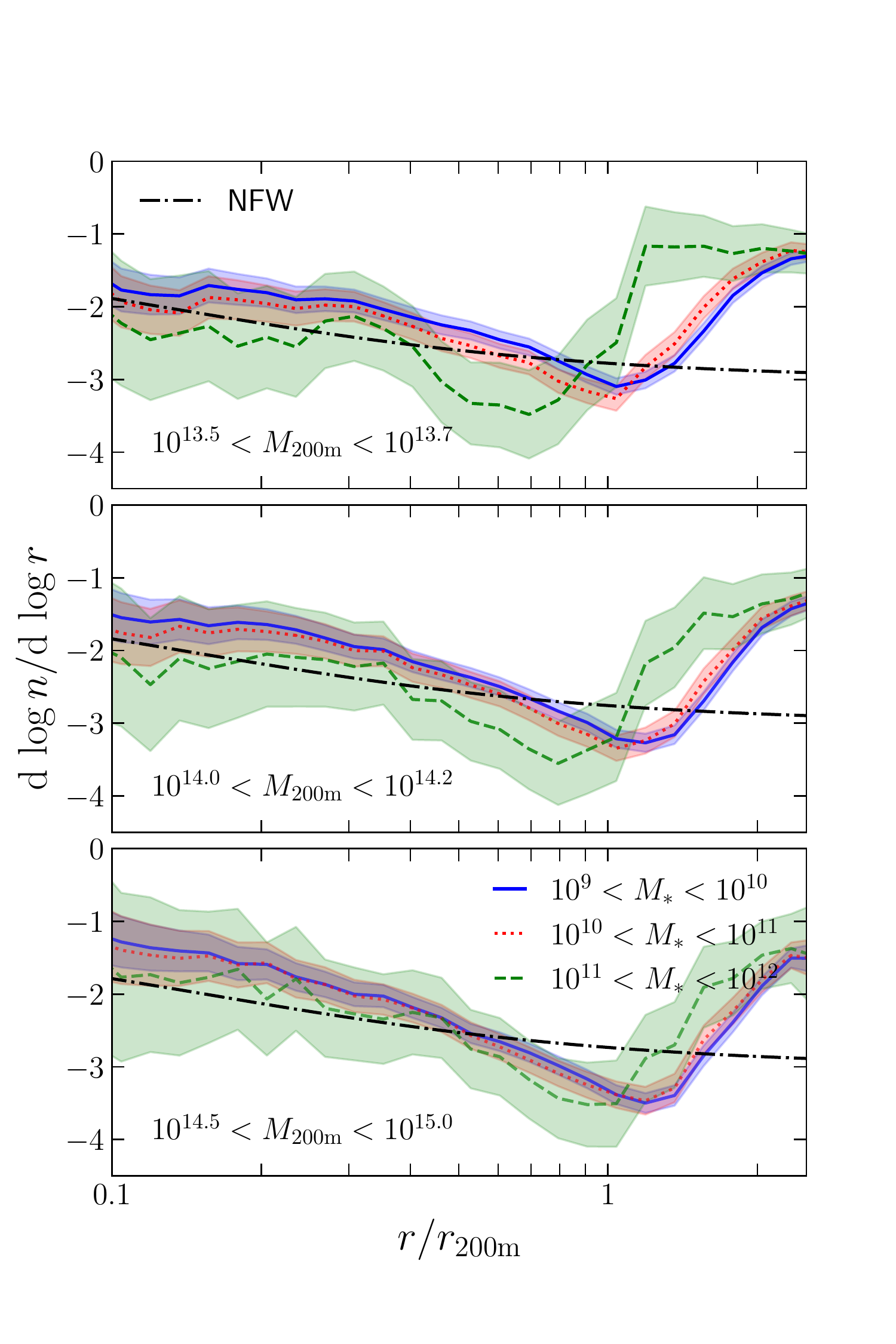}
    \caption{The logarithmic slope of the number-density profile of galaxies as a function of the normalized radius, $r/r_{\rm 200m}$, at $z=0$. Different panels show the results for different central halo mass ranges in the unit of $M_{\odot}/h$. Each panel shows the profiles for different satellite stellar mass bins in the unit of $M_{\odot}$ using different colors and line types. The slope follows the NFW profile (black dot-dashed line) and steepens near $\r200m$. The error bars on the slope are propagated from the error on the mean of the average profiles. The errors in the largest stellar mass bin are quite large due to the lack of satellite galaxies with a large mass.}
    \label{fig:number_density_slope}
\end{figure}

\Cref{fig:number_density_slope} shows the logarithmic slopes of the average number-density profiles of galaxies in halos with different masses. Here we select galaxies by their stellar mass as indicated in the plot. The shaded regions present the errors on the slope propagated from the errors on the mean of the profiles. Overall, the profiles tend to follow the NFW profile of the dark matter \citep{NFW}, using the mass--concentration relation from \citet{Uchuu}.

However, there are important differences. The steepening feature at $r\sim r_{\rm 200m}$ and the bump at $r\sim 2r_{\rm 200m}$ is the result of a pile-up of backsplash galaxies which are at their second turnaround after infalling into the central halo for the first time. 
  
The profiles do not show a significant dependence on the central halo mass once the radius is normalized by $\r200m$. However, the slope increases with the increasing stellar mass of galaxies. In addition, the radius, where the density slope steepens, is smaller for galaxies with larger stellar mass. This is likely due to the dynamical friction, which is stronger for higher mass galaxies, reducing their kinetic energy and, thus, the second turnaround radius \citep{adhikari_etal2014}.

\subsection{Galaxy Quenching}
\label{sec:quench}

\begin{figure}
    \centering
    \includegraphics[width=0.49\textwidth]{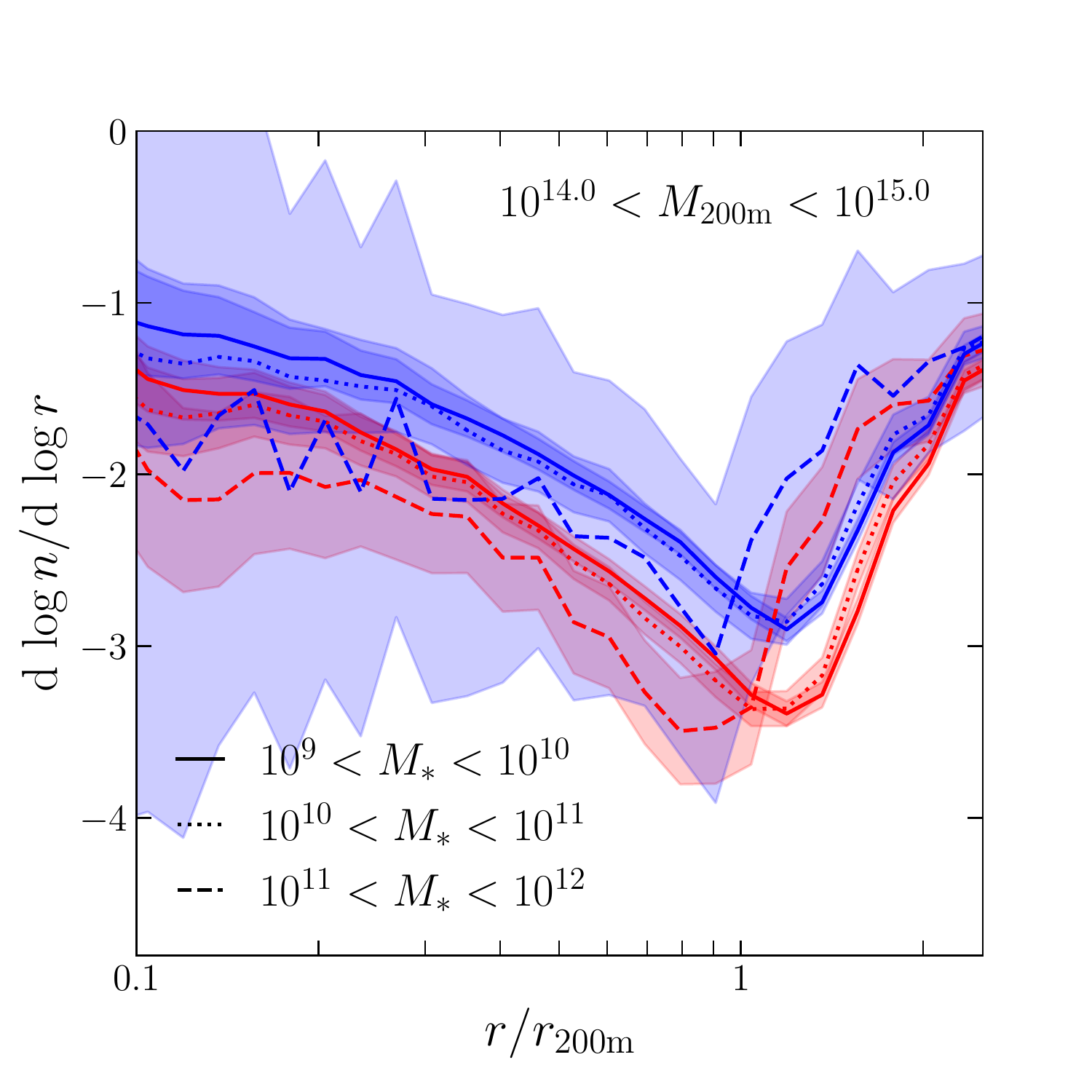}
    \caption{The slope of the number-density profiles of galaxies as a function of radius for different satellite stellar masses in units of $M_{\odot}$ (different line styles) and for different types of galaxies (red color for red quenched galaxies, blue color for blue star-forming galaxies) at $z=0$. Red galaxies exhibit steeper slopes than blue galaxies, but galaxies of the same mass have the same splashback radius regardless of the color of the galaxies. The error bars on the slope are propagated from the error on the mean of the average profiles. The errors in the largest stellar mass bin for blue galaxies are quite large because there are very few satellite galaxies with large masses, which are also star-forming.}
    \label{fig:number_density_red}
\end{figure}

\begin{figure}
    \centering
    \includegraphics[width=0.49\textwidth]{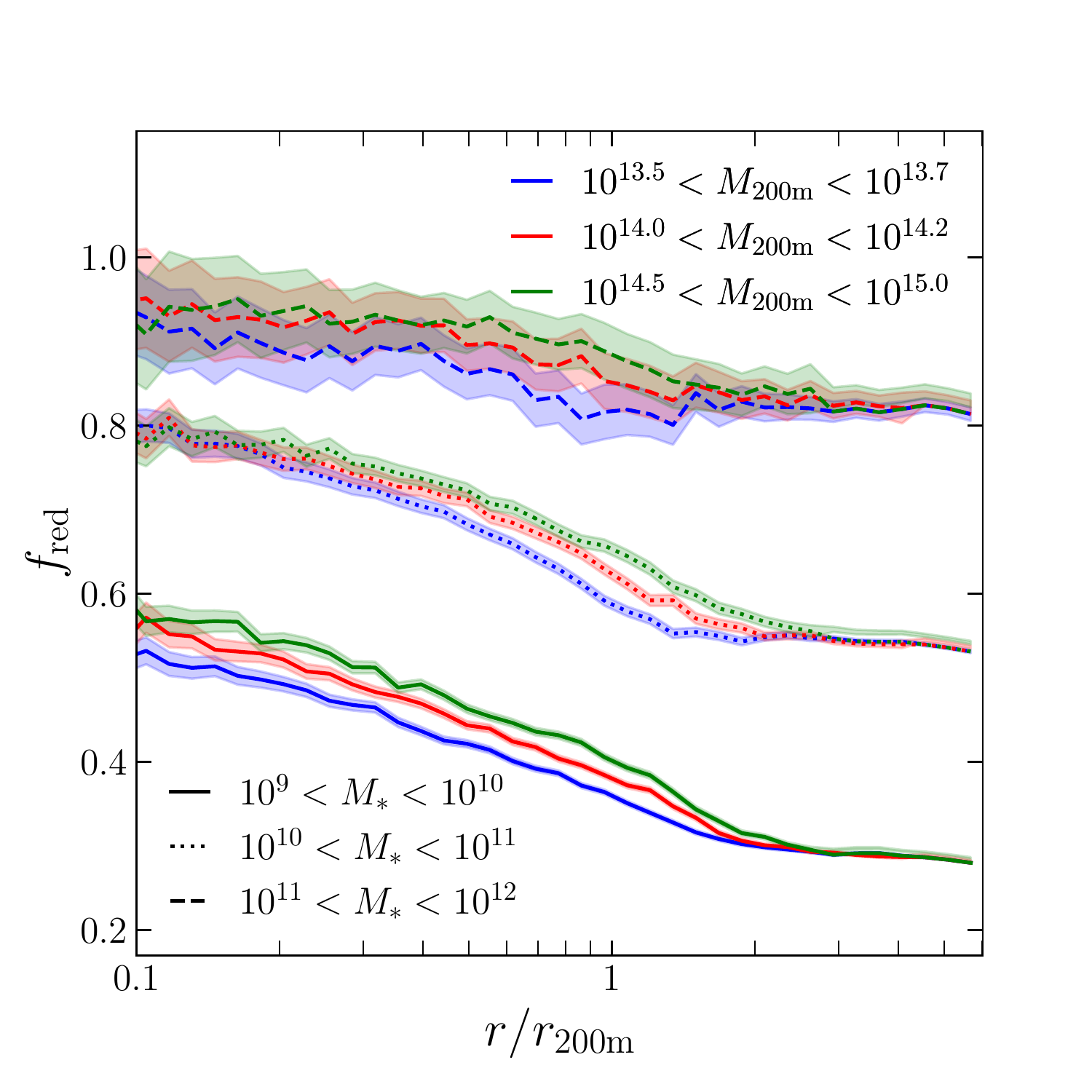}
    \caption{Quenched fraction of galaxies at $z=0$ as a function of cluster-centric distance for different central halo masses in units of $h^{-1}M_{\odot}$ (different colors) and different satellite stellar mass in units of $M_{\odot}$ (different linestyles).  The larger the stellar mass, the larger the quenched fraction. The quenched fraction increases with dark matter halo mass for similar stellar mass. The quenched fraction decreases monotonically with the halo radius and asymptotically approaches the global quenched fraction at $r\sim 2r_{\rm 200m}$.}
    \label{fig:fred}
\end{figure}
 
\Cref{fig:number_density_red} shows that the slope of the number-density profile of galaxies depends on the galaxy color. The slope of red galaxies is steeper than for blue galaxies at all radii, including the steepening feature around the splashback radius. Red galaxies, on average, have spent a long time in the dark matter halo, experiencing tidal stripping and dynamical friction for a more extended period. Hence, the difference in the slope between high and low stellar mass galaxies also appears for red and blue galaxies. Note, however, that the radius at which the steepest density slope occurs is similar for galaxies of different star formation rates at the same stellar mass. 

\Cref{fig:fred} shows the fraction of quenched galaxies as a function of halo radius around galaxy clusters and groups. The fraction of red galaxies decreases monotonically from the inner region to the outskirt. It asymptotically approaches the global quenched fraction at the large cluster-centric radius. In the cluster outskirts, the main culprit of quenching in field galaxies is baryonic feedback, whose effect is more pronounced in higher mass halos, explaining why there are more red galaxies among galaxies with larger stellar masses. The quenched fraction beyond $3r_{\rm 200m}$ approaches the global quenched fraction, indicating the limits of the impact of the clusters and groups. Galaxies in a high-density environment (such as galaxy clusters) tend to quench due to strangulation by ram pressure and tidal stripping and starvation due to lack of cold gas inflow or feedback effects \citep{Cortese2021}. This creates an increasing gradient in the fraction of red galaxies from the outer to the inner region of the dark matter halo. The fraction deviates from the global fraction starting around $r=(1.5-2)\r200m$ and increases as the separation becomes smaller. This behavior has been reported in the observed color gradients of the galaxy clusters \citep[e.g.,][]{Wetzel2014, Adhikari2021}. The excess fraction of red galaxies outside $\r200m$ and splashback radius is due to backsplash galaxies: galaxies that have had pericentric passage inside the cluster and experience quenching can backsplash and reach beyond virial radius, extending out to $\sim 2\r200m$ \citep{Balogh2000,Mamon2004,ludlow2009,wang2009,Aung2021}.

The global quenched fraction at large radii and the quenched fraction in the cluster is a strong function of stellar mass. At large radii, only $30\%$ of the galaxies with $M_*=10^{9}-10^{10}M_\odot$ are quenched, while $80\%$ of the galaxies with $M_*=10^{11}-10^{12}M_\odot$ are quenched. The quenched fraction for galaxies increases from $r=2\r200m$ to $r=0.1\r200m$ by 0.2-0.3 for all stellar mass bins, and at $r=0.1\r200m$, $>90\%$ of galaxies with $M_*=10^{11}-10^{12}M_\odot$ are quenched, while only $60\%$ of the galaxies with $M_*=10^{9}-10^{10}M_\odot$ are quenched. The quenched fraction has a very mild dependence on the halo mass. This behavior is qualitatively similar to the observational measurements in \citet{Wetzel12}, where the overall quenched fraction in the cluster is a strong function of stellar mass. The excess quenching in the cluster environment (i.e., the difference between the quenched fraction in the cluster relative to the field) is independent of the stellar mass of galaxies. We also found that the larger the halo mass, the larger the quenched fraction is for the same stellar mass galaxies, indicating stronger environmental effects. However, the fraction of quenched galaxies changes by less than 0.05 in the halo mass range probed.

Note that our mock catalogue reproduces the expected features of galaxy quenching despite the lack of explicit models of environmental effects. \UM does not consider whether a galaxy is near or inside clusters and groups but only considers the changes in the halo's $v_{\rm peak}$. If $v_{\rm peak}$ of the halo stops growing, i.e., the halo stops accreting mass, the galaxy will have a lower star formation rate. The fact that the catalogue produces the observed environmental effect indicates that the halo's most influencing factor in determining the star formation rates of the galaxies is the mass accretion rate.

\subsection{Projected Galaxy Profiles around Galaxy Clusters and Groups}
\label{sec:proj}

\begin{figure*}
    \centering
    \includegraphics[width=0.99\textwidth]{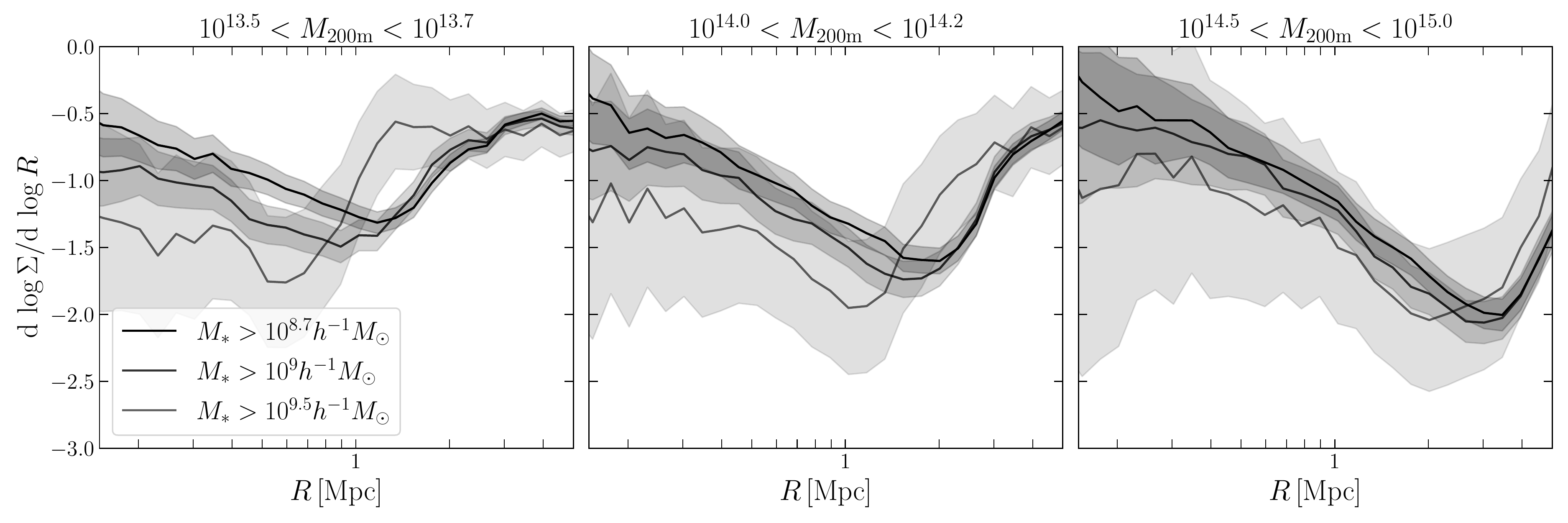}
    \caption{The logarithmic slope of the number density of galaxies as a function of projected radius for different halo mass (different panels) and different galaxy mass (different line styles). The profiles follow the same qualitative feature as 3D profiles in \Cref{fig:number_density}.}
    \label{fig:number_density_2d_red}
\end{figure*}

\begin{figure*}
    \centering
    \includegraphics[width=0.99\textwidth]{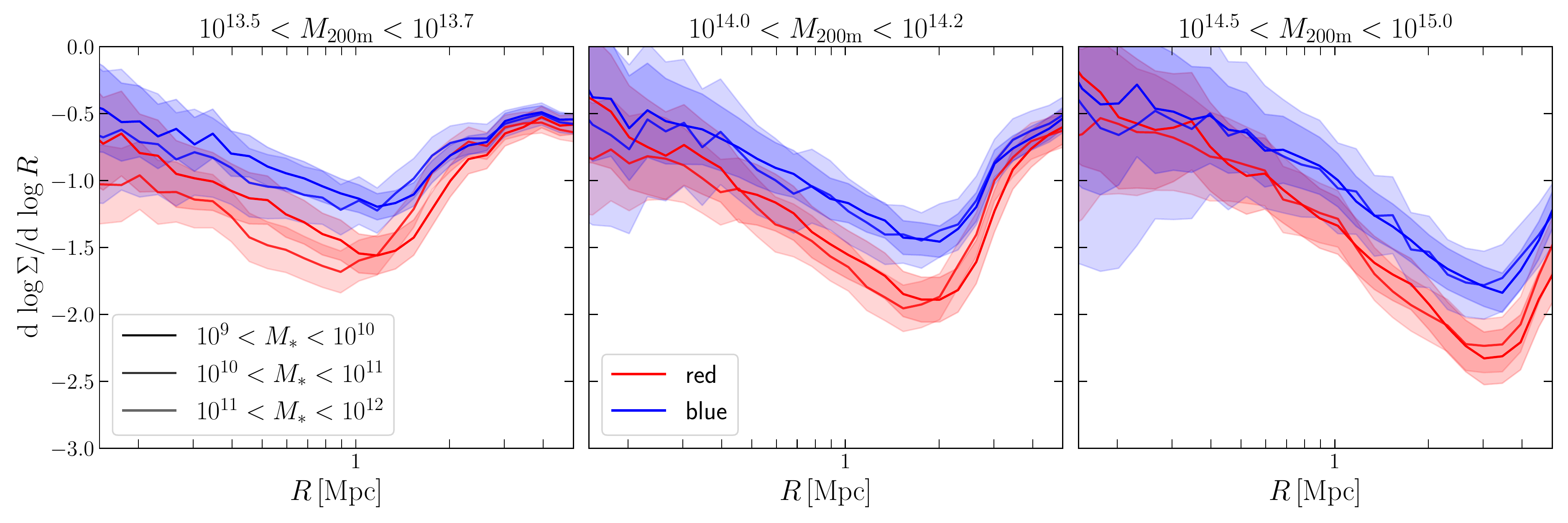}
    \caption{The logarithmic slope of the number density of galaxies as a function of projected radius for different colors of galaxies. The profiles follow the same qualitative feature as 3D profiles in \Cref{fig:number_density_red}, where the slope of the number density profile of blue galaxies is shallower than red galaxies.}
    \label{fig:number_density_2d}
\end{figure*}

\begin{figure*}
    \centering
    \includegraphics[width=0.99\textwidth]{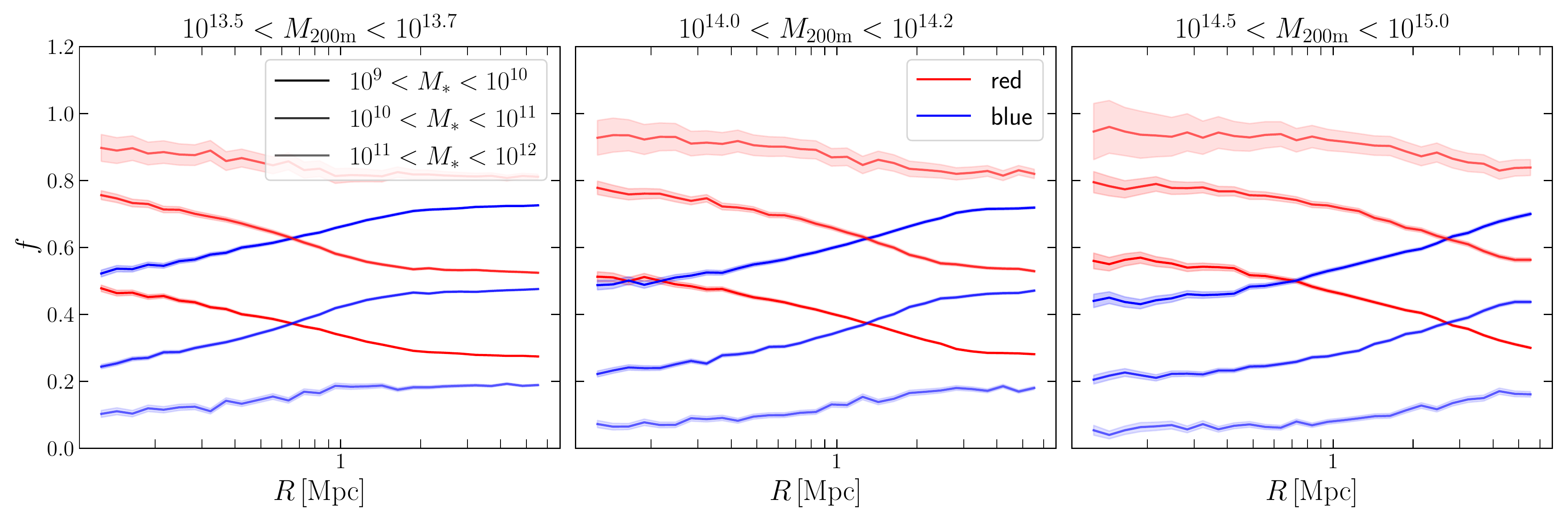}  
    \caption{Fraction of red and blue galaxies as a function of projected radius. The projected profile again follows the same behavior as 3D profiles in \Cref{fig:fred_2d} as quenched fraction decreases as the galaxy is farther away from the cluster and asymptotes to the global quenched fraction at large radii.}
    \label{fig:fred_2d}
\end{figure*}

Finally, building on the physical intuition based on the three-dimensional profiles, we investigate two-dimensional projected profiles of galaxy properties to connect with observations. 
We generate a mock catalogue by projecting the simulation box along the $z$ axis. We select galaxies with the line-of-sight velocity with respect to the clusters within $|\Delta v|<1500\kms$. 

\Cref{fig:number_density_2d} shows the logarithmic slope of the projected number density profile of galaxies similar to \Cref{fig:number_density} but as a function of projected radius. The profiles show similar qualitative features as 3D profiles, where the slope of the density profile decreases and steepens. In contrast, the splashback radius decreases as the stellar mass of the galaxies increases. The slope is offset by about $-1$ from 3D profiles at most radii due to the integral along line-of-sight.

\Cref{fig:number_density_2d_red} shows the logarithmic slope of the projected number density profile but for different galaxy colors, separately for red and blue galaxies. Similar to \Cref{fig:number_density_red}, red galaxies exhibit an overall steeper slope but are offset by about $-1$.
\citet{Adhikari2021} reports that in observed galaxies in 2D, the number density profile of red galaxies is steeper than that of blue galaxies similar to our findings in 3D. The splashback radius of red galaxies is also smaller than that of blue galaxies, which is not found in our studies. This is likely caused by different mass selections in red vs. blue galaxies.

\Cref{fig:fred_2d} shows the quenched fraction as a function of the projected radius. Similar to the number density profile, the 2D profiles show a qualitative feature similar to 3D profiles where the quenched fraction decreases as a function of the projected radius and asymptotes to the global fraction at a large radius.

\section{Conclusions}
\label{sec:conc}

In this work, we presented the Uchuu-UM mock galaxy catalogue for a large volume with a box size of $2\hinv\Gpc$ on the side and mass range $10^{10}<M_{\rm halo}/M_{\odot}<5\times10^{15}$, including over $10^{4}$ cluster-size halos with galaxies down to $ M_*\sim5\times 10^8\Msun$. Key features of the Uchuu-UM  mock galaxy catalogue are summarized below: 

\begin{itemize}
    \item The catalogue reproduces the observed data compiled by \citet{Behroozi19} including projected correlation function and quenched fractions (see \Cref{fig:correlation,fig:quenched_fraction}). 
    \item The stellar mass function (\Cref{fig:smf}) agrees with the previous \UM datasets. Both the stellar mass function and satellite stellar mass function for satellite galaxies at given dark matter halo mass show completeness above $\sim 5\times10^{8}\Msun$ (\Cref{fig:csmf}). Compared to the MDPL2 simulation, the Uchuu-UM catalogue includes significantly more massive galaxies hosted by large-mass dark matter halos.
	\item Overall, the number-density profile of galaxies follows the NFW profile of dark matter halos. The profile becomes steeper around $r_{\rm 200m}$, where the slope is $d\log n/d\log r \sim -3$. At larger radii, the profile flattens again in the region of the splashback radius $\sim (2-3)r_{\rm 200m}$. The density profile tends to be steeper for larger stellar masses.
	\item It is important to include orphan galaxies. If ignored, the profile quickly flattens in the inner region as several subhalos are artificially disrupted. (\Cref{fig:number_density,fig:number_density_slope})
    \item The number-density profile of galaxies depends on the color of galaxies, with red galaxies having steeper slopes at all radii when compared to blue galaxies. The red galaxies tend to steepen more at the splashback radius, while the steepening feature is weaker in blue galaxies (\Cref{fig:number_density_red}).
	\item The fraction of quenched galaxies in and around cluster-size halos shows a remarkably complex behaviour. As expected, there are more quenched galaxies in the inner regions of the halos. However, at a fixed stellar mass the dependence on cluster mass is remarkably weak: at $r=0.1r_{\rm 200m}$ and for $M_*=10^{10}-10^{11}M_\odot$ the quenched fraction is $0.55$ for $M_{200m}\sim 3\times 10^{13}M_\odot$. It is nearly the same $0.58$ for $M_{200m}\sim 5\times 10^{14}M_\odot$. 
	\item We find by far the strongest trend of the quenched fraction on the stellar mass. In central cluster regions $\sim 90\%$ of massive galaxies $M_*>10^{11}M_\odot$ are quenched while only $\sim 50\%$ are quenched for $M_*<10^{10}M_\odot$.  The quenched fraction then asymptotically approaches the global quenched fraction at large radii. The fraction increases toward the inner regions of the cluster-centric radius, indicating quenching in cluster environments (\Cref{fig:fred}).
	\item The projected number density (\Cref{fig:number_density_2d}) as well as quenched fraction (\Cref{fig:fred_2d}) follow similar qualitative feature as 3D profiles, but with the density slope offset by about $-1$.
\end{itemize}

The Uchuu-UM galaxy catalogue presented here represents a step forward in modeling ongoing and upcoming large galaxy surveys. Recent advances in hydrodynamic cosmological simulations provide an alternative approach for modeling galaxy formation and evolution \citep[e.g.,][]{Hirschmann2014MGTM,Eagle,McCarthy2017BAHAMAS,Pillepich2018FirstGalaxies,The300}. In the future, we plan to compare the predictions of the Uchuu-UM galaxy catalogue to a plethora of observables (such as the radial distribution of cluster galaxies, splashback radius, and galaxy quenching) in hydrodynamic cosmological simulations and observations. 

\section*{Acknowledgements}

We thank the Instituto de Astrof\'isica de Andaluc\'ia (IAA-CSIC), Centro de Supercomputaci\'n de Galicia (CESGA), and the Spanish academic and research network (RedIRIS) in Spain for hosting Uchuu DR1 in the \SU{} site (\urluchuu) for cosmological simulations. We are grateful to Antonio Fuentes, and his team at RedIRIS, for providing the skun@IAA\_RedIRIS server that hosts Uchuu DR1 and DR2 through \SU{} with a fast download speed through their RedIRIS High Speed Data Transfer Service; to the staff at CESGA for developing the Uchuu-BigData platform and providing the computer resources; and the staff at the IAA-CSIC computer department for proving support and managing the skun@IAA\_RedIRIS server.

The Uchuu simulations were carried out on Aterui II supercomputer at the Center for Computational Astrophysics, CfCA, of the National Astronomical Observatory of Japan, and the K computer at the RIKEN Advanced Institute for Computational Science (Proposal numbers hp180180, hp190161). The Uchuu-UM galaxy catalogue data release has used the skun6@IAA facility managed by the IAA-CSIC in Spain; this equipment was funded by the Spanish MICINN EU-FEDER infrastructure grant EQC2018-004366-P. The MICINN grant AYA2014-60641-C2-1-P funded the skun@IAA\_RedIRIS server.

DN and HA acknowledge support from Yale University and the Yale Center for Research Computing facilities and staff. HA also acknowledges support from the Zuckerman Postdoctoral Scholars Program while completing this work. 
T.I. has been supported by IAAR Research Support Program at Chiba University Japan, 
MEXT/JSPS KAKENHI (Grant Number JP19KK0344, JP21F51024, and JP21H01122), 
MEXT as ``Program for Promoting Researches on the Supercomputer Fugaku'' (JPMXP1020200109), and JICFuS. FP, AK, EP, JR thank the support of the Spanish Ministry of Science and Innovation funding grant PGC2018-101931-B-I00. 

Numerical analyses were partially carried out using the following packages, \textsc{numpy} \citep{numpy}, 
\textsc{scipy} \citep{scipy}, \textsc{h5py}\citep{h5py},
\textsc{matplotlib} \citep{matplotlib}, 
\textsc{halotools} \citep{Hearin2017}, and \textsc{colossus} \citep{Diemer2018}.

%%%%%%%%%%%%%%%%%%%%%%%%%%%%%%%%%%%%%%%%%%%%%%%%%%
\section*{Data Availability}
The Uchuu-UM galaxy data is publicly available on \urluchuu \, along with the Uchuu simulation data release. Non-public data are available upon request. 
Under the UchuuProject on Github~\footnote{\url{https://github.com/uchuuproject}}, 
we provide a collection of general tools for the community, 
codes / scripts / tutorials and Jupyter notebooks for analysing the Uchuu data.


\bibliographystyle{mnras}
\bibliography{ref} % if your bibtex file is called example.bib

%%%%%%%%%%%%%%%%%%%%%%%%%%%%%%%%%%%%%%%%%%%%%%%%%%

% Don't change these lines
\bsp	% typesetting comment
\label{lastpage}
\end{document}